\documentclass[twocolumn,pacs,aps,prl,floatfix,superscriptaddress]{revtex4}
\usepackage{amsmath,amssymb}
\usepackage{graphicx,float}
\usepackage{times,txfonts}
\usepackage{color}
\usepackage{multirow}
\usepackage{bbold}
\usepackage{color}
\usepackage{soul}
\usepackage{hyperref}
\usepackage{times,txfonts}
\usepackage{natbib}

\newcommand{\braket}[2]{\left\langle {#1{\left| \vphantom{#1 #2} \right.} #2} \right\rangle}
\newcommand{\qo}[1]{``#1''}

\renewcommand{\epsilon}{\varepsilon}
\renewcommand{\phi}{\varphi}

\usepackage{sistyle}
\usepackage{soul}
\definecolor{lightblue}{RGB}{185,210,248}
\sethlcolor{lightblue}

\begin{document}
\title{Holographic generation of highly twisted electron beams}
\author{Vincenzo Grillo}
\affiliation{CNR-Istituto Nanoscienze, Centro S3, Via G Campi 213/a, I-41125 Modena, Italy}
\affiliation{CNR-IMEM Parco Area delle Scienze 37/A, I-43124 Parma, Italy}
\author{Gian Carlo Gazzadi}
\affiliation{CNR-Istituto Nanoscienze, Centro S3, Via G Campi 213/a, I-41125 Modena, Italy}
\author{Erfan Mafakheri}
\affiliation{CNR-Istituto Nanoscienze, Centro S3, Via G Campi 213/a, I-41125 Modena, Italy}
\affiliation{Dipartimento FIM, Universit\'a di Modena e Reggio Emilia, Via G. Campi 213/a, I-41125 Modena, Italy}
\author{Stefano Frabboni}
\affiliation{CNR-Istituto Nanoscienze, Centro S3, Via G Campi 213/a, I-41125 Modena, Italy}
\affiliation{Dipartimento FIM, Universit\'a di Modena e Reggio Emilia, Via G. Campi 213/a, I-41125 Modena, Italy}
\author{Ebrahim Karimi}
\email{ekarimi@uottawa.ca}
\affiliation{Department of Physics, University of Ottawa, 25 Templeton, Ottawa, Ontario, K1N 6N5 Canada}
\author{Robert W Boyd}
\affiliation{Department of Physics, University of Ottawa, 25 Templeton, Ottawa, Ontario, K1N 6N5 Canada}
\affiliation{Institute of Optics, University of Rochester, Rochester, New York, 14627, USA}
\begin{abstract}
Free electrons can possess an intrinsic orbital angular momentum, similar to those in an electron cloud, upon free-space propagation. The wavefront corresponding to the electron's wavefunction forms a helical structure with a number of twists given by the \emph{angular speed}. Beams with a high number of twists are of particular interest because they carry a high magnetic moment about the propagation axis. Among several different techniques, electron holography seems to be a promising approach to shape a \emph{conventional} electron beam into a helical form with large values of angular momentum. Here, we propose and manufacture a nano-fabricated phase hologram for generating a beam of this kind with an orbital angular momentum up to 200$\hbar$. Based on a novel technique the value of orbital angular momentum of the generated beam are measured, then compared with simulations. Our work, apart from the technological achievements, may lead to a way of generating electron beams with a high quanta of magnetic moment along the propagation direction, and thus may be used in the study of the magnetic properties of materials and for manipulating nano-particles.
\end{abstract}
\maketitle

Almost a century ago Rutherford and Bohr proposed a model, the so-called \emph{Bohr model}, to describe the structure of atoms in which model atoms consist of a positive nucleus surrounded by orbiting electrons~\cite{rutherford:11,bohr:13}. Even in this semi-classical model, orbiting electrons possess a quantized orbital motion, i.e. orbital angular momentum (OAM). This quantization, indeed, lies at the heart of the rotationally symmetric nature of the atom. However, it took quite a long time to theoretically predict and experimentally demonstrate that free electrons can also carry a quantized OAM value upon free-space propagation~\cite{bliokh:07,uchida:10,verbeeck:10}. The wavefront of electrons carrying a quantized OAM forms a helical shape $\exp{(im\phi)}$ with an integer winding index $m$, where $\phi$ is the azimuthal angle in polar coordinates. A free electron with such a helical phasefront possesses an OAM value of $m\hbar$ along the propagation direction, and has a magnetic moment $\mu_{\text{OAM}}=m\mu_B$ oriented along the beam axis with a polarity that depends on the sign of $m$. $\mu_B=e\hbar/(2m_e)$ is the Bohr magneton of the electron, $\hbar$ is the Planck constant, $e$ and $m_e$ are the electron charge and rest mass, respectively. This magnetic moment, unlike the spin Bohr magneton, in principle is unbounded and can be large if desired. Nonetheless, it is bounded by the accuracy of phase modulation and the numerical aperture of the electron optics~\cite{mcmorran:11}. The spatial density distribution of these electrons in the transverse plane -- orthogonal to propagation direction -- appears to be a doughnut shape, because a helical phase is undefined at the origin. Moreover, the current density associated with the wavefunction of \qo{twisted} electrons circulates about the origin; thus, these beams are also called electron vortex beams (EVBs). Twisted electron beam (EBs) possess a novel \emph{magnetic moment} $\mu_{\text{OAM}}$ along the propagation axis, and thus found immediate applications in the study of materials~\cite{rusz:14,verbeeck:14}. Among those applications, manipulating nano-particles~\cite{lloyd:13} and measuring magnetic dichroism~\cite{lloyd:12b} are primary examples. In the latter case, the magnetic moment of EVBs, in addition to spin-$\frac{1}{2}$ of the electron is coupled into internal dynamics of atoms~\cite{verbeeck:13}. Several different methods analogous to optical counterparts such as spiral-phase plates~\cite{uchida:10}, fork-holograms~\cite{verbeeck:10,mcmorran:11,grillo:14}, astigmatic mode convertors~\cite{schattschneider:12}, spin-to-OAM~\cite{karimi:12,karimi:14}, and tuning of multipolar aberration corrector~\cite{petersen:13,clark:13} have been proposed to generate twisted electrons, some have been experimentally verified. 

In this Letter, we report the generation of twisted EBs with an OAM value of $200\hbar$ --  the highest electron OAM quanta up to now. Achieving an EB with high number of twists provides the possibility of exploring the transition between the quantum and classical regime of electromagnetic radiations inside a medium, where the spin-induced effects are diminished~\cite{ivanov:13}. This transition is particularly interesting in view of observing phenomena such as polarization radiation, which has never been experimentally observed. The generated beam is manipulated by a nano-fabricated phase hologram with 200 $2\pi$-phase variations around the origin. Such a high phase-variation around the origin is beyond the hologram's spatial-fringe resolution. Thus, we examine the OAM-purity of the generated beam for two different cases where the centre of the hologram is excluded or included. The results reveal that the generated beam carries a specific \emph{OAM spectrum} determined by the resolution of phase modulation of the hologram. Thus, for this special case the region beyond the hologram's spatial resolution must be excluded. Finally, we analyze the radial distribution of the generated beam that is relevant to study electron-magnetic interactions~\cite{bliokh:12}.

Let us now briefly discuss the holographic approach of generating an EB with a helical phasefront of $\braket{\mathbf r}{\psi_m}=\exp{(im\varphi)}$. Such a beam can be generated in different ways; \emph{(i)} a pure phase object (spiral phase-plate) with linear phase scaling, \emph{(ii)} amplitude mask with a dislocation at the centre and \emph{(iii)} a pure phase mask (PM) with limited phase mapping in the interval of $[a,b]$, where $a,b\in\{-\pi,\ldots,\pi\}$ (examples are reported in Refs.~\cite{voloch:13,grillo:14a,harvey:14}). The latter case gives a more realistic way to generate an EB with a desired shape, since a carrier can be used to \qo{sift} the desired beam from the reference beam. The shape of the PM used to generate a twisted EB is identical to the amplitude mask, but entirely transparent to the electrons. Thus, almost all electrons travel through the mask without being absorbed. Nonetheless, electron-electron elastic scattering introduced by the mask changes the phase of the electron wavepacket in the transverse plane. This phase alteration is proportional to the mean inner potential of the material, $V_{mip}$~\cite{reimer:08}. A PM with a thickness of $t(x,y)$ varying in the transverse $x-y$ plane introduces a coordinate-dependent phase change of $\Delta \chi(x,y)$, given by:
\begin{align}\label{eq:phase}
	\Delta \chi=\frac{2\pi e}{\lambda}\,\frac{{\cal E}+{\cal E}_0}{{\cal E}({\cal E}+2{\cal E}_0)}\,V_{mip}\,t(x,y),
\end{align}
where $\lambda$ is the de Broglie wavelength of the electron, ${\cal E}$ and ${\cal E}_0$ are the kinetic and the rest energy of the electrons, respectively~\cite{muller:05}. The thickness function $t(x,y)$ determines the induced phase alteration, and consequently the efficiency of the generated beam. 

\begin{figure}[t]
   \centering
   \includegraphics[width=0.38\textwidth]{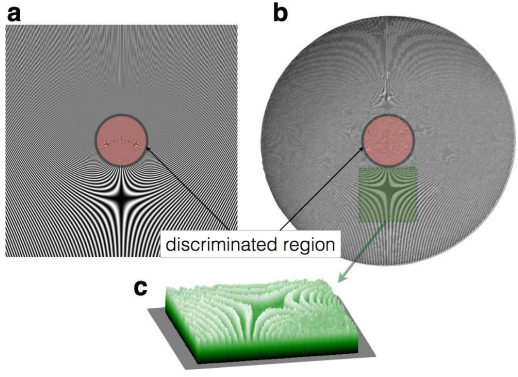} 
   \caption{(online color) ({\bf a}) Designed CGH for generating an EVB with 200 twists. This CGH imprints a specific phase delay between $[-\pi,\pi]$ onto the impinging beam. ({\bf b}) SEM image of the nano-fabricated phase hologram shown in ({\bf a}). The hologram forms a pitch-fork shape with 200 dislocations at the centre. Due to high oscillation of the optical phase around the origin, shown in pink circles, the carrier changes more rapidly in a specific region where the phase alteration goes beyond the spatial resolution. ({\bf c}) Thickness map of the shaded region of the fabricated hologram (see SM for more details~\cite{SM}).}
   \label{fig:fig1}
\end{figure}

In this work, we focus on generating an EB with a high number of twists. Thus, at a cost of efficiency, a PM with a sinusoidal modulation is used to generate EBs, because it provides better control on the mask structure. Therefore, we implement a sinusoidal modulation, i.e. $t(x,y)=t_0\left(1+\cos{\left(m\,\phi+2\pi x/\Lambda\right)}\right)/2$, where $t_0$ is the modulation depth, $\Lambda$ is the grating period and $\phi=\arctan{(y/x)}$ with $x$ and $y$ being cartesian coordinates (see Supplementary Material (SM) for more details~\cite{SM}). The structure of the PM and the scanning electron microscope (SEM) image of the fabricated hologram are shown in Fig.~\ref{fig:fig1}-({\bf a}) and ({\bf b}), respectively (fine structure of the nano-fabricated computer generated holograms (CGHs) are shown in SM~\cite{SM}).
%
%
The nano-fabricated CGH, shown in Fig.~\ref{fig:fig1}-({\bf b}), is generated with the same approach reported in Ref.~\cite{grillo:14, grillo:14a}, where a Ga-ion focused ion beam (FIB) is used to mill a silicon-nitride membrane. The thickness map of the fabricated hologram measured by energy loss mapping is shown in Fig.~\ref{fig:fig1}-({\bf c}). The thickness of the hologram varies in the range of 120-180~nm, and the modulation depth is $t_0=30$~nm. This provides enough phase change to tailor the phase of the electron wavepacket while for such large thickness the modulation in absorption is negligible.

As can be seen in Fig.~\ref{fig:fig1}, there exists a region (shown in pink circles in Fig.~\ref{fig:fig1}- ({\bf a}) and ({\bf b})) where the phase alteration is almost undefined, due to the presence of a phase singularity. This \emph{dead region} can be blocked by an obstacle, as was proposed originally in Ref.~\cite{mcmorran:11}, because the portion of the beam which passes through it is not affected by the mask. Nevertheless, as will be shown later, it affects the beam quality. Thus, we decided to compare and analyze the OAM spectrum of the generated beam for two different cases; when this dead central region is kept, and the case where the central region is obstructed.

\begin{figure}[t]
   \centering
   \includegraphics[width=0.42\textwidth]{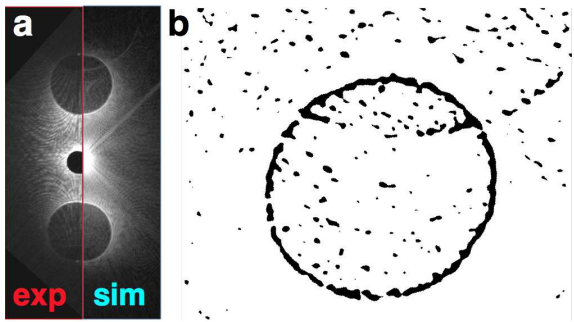} 
   \caption{(online color) ({\bf a}) Experimentally (exp) observed and simulated (sim) Fraunhofer diffraction of a conventional EB from the CGH of Fig.~\ref{fig:fig1}-({\bf b}). The upper and lower bright doughnuts are the $\pm1$-orders of diffractions. In order to increase the visibility of the first-order of diffraction, the central beam is blocked by a beam stop. The hologram is symmetric, and thus both first orders of diffraction possess the same OAM value, nonetheless having an opposite sign. The distribution of the electrons that possess a phase singularity at the upper side ({\bf a}) is calculated and shown in ({\bf b}). The singularity tends to concentrate in a ring at the periphery of the dark region. This is achieved by measuring phase variation about an infinitesimal region close to each pixel at the upper diffraction area. As can be seen, the residual of second-order of diffraction with an OAM value of 400$\hbar$ (upper doughnut) is partially superimposed with the beam of first-order. This lies in the fact that the diffraction angle of the electrons leaving the hologram is smaller than the divergence angle of the EB. Similar results are achieved for the hologram that its central region, shown in Fig.~\ref{fig:fig1}, is blocked.}
   \label{fig:fig2}
\end{figure}

We illuminated the hologram with a Schottky Field emitter generated beam (a relatively coherent EB), in a JEOL 2200Fx transmission electron microscope (TEM) with a convergence angle below 0.3~$\mu$rad and a central energy of ${\cal E}=200$~keV. This corresponded to a de Broglie wavelength of $\lambda=2.5$~pm. The hologram, Fig.~\ref{fig:fig1}-({\bf b}), is inserted in the specimen position of the electron microscope. The EVB is generated in the low-magnification Lorentz mode as explained in~\cite{grillo:14}. The distribution of the electrons at the Fraunhofer plane is recorded, which provides a maximum separation between all orders of diffractions. Figure~\ref{fig:fig2}-({\bf a}) shows the distribution of the electrons for first orders of diffraction, where the central beam is blocked by a beam stop. The first-order of diffraction is zoomed in and shown in the map of Fig.~\ref{fig:fig2}-({\bf b}). The generated EB is not pure, since there is residual of the beam at the second-order of diffraction with OAM value of $400\hbar$ that overlaps with the first-order beam. However, the generated beam at the first-order of diffraction is a quasi-coherent superposition of OAM states peaked about $200\hbar$ with a different width and shape for two different types of holograms discussed above. The generated EVB, at the first-order, can then be spatially filtered by means of an aperture. It is noteworthy to mention that the largest value of electron OAM ever reported~\cite{mcmorran:11} is created at the fourth-order of diffraction with a very low efficiency. Conversely, in our case the generated beam is positioned at the first-order of diffraction, which thanks to efficiency of the PM it is much more intense ($\gtrsim 30$ times brighter).\newline
%
%
\begin{figure}[b]
   \centering
   \includegraphics[width=0.35\textwidth]{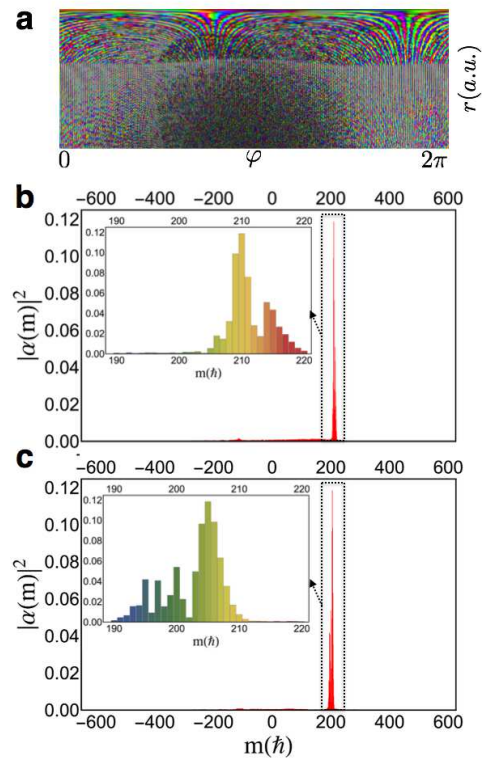} 
   \caption{(online color) ({\bf a}) Calculated phase distribution of the electron wavefunction around the first-order of diffraction. This is calculated by Fourier transforming the experimentally determined transfer function of the hologram from a \emph{measured map} of the hologram thickness (see Fig.~\ref{fig:fig1}-({\bf c})). The phase distribution is shown in polar coordinates, where the phase value is represented in a hue-color. The OAM spectrum of the generated beam at the first-order of diffraction from a the hologram with and without the central \emph{dead} region are shown in ({\bf b}) and ({\bf c}), respectively. This is measured by performing a one-dimensional Fourier transform of the azimuth coordinate. Due to imperfections at the origin and low spatial frequency for the case of ({\bf b}), the beam is not pure and the value of OAM is delocalized centred around OAM of $(210.95\pm0.030)\hbar$. While for the hologram with the excluded central dead region ({\bf c}) the OAM spectrum central peak is about $(202.75\pm0.032)\hbar$. Indeed, the OAM spectrum, for both cases, moves to higher values mainly due to a small contribution from the second-order of diffraction which carries OAM of $400\hbar$ (see Fig.~\ref{fig:fig2}-({\bf b})). Nevertheless, the hologram with the excluded central dead region ({\bf c}) gives much better results.}
   \label{fig:fig3}
\end{figure}
%
As the next step, the quality of generated highly-twisted EBs for these two different holograms is analyzed. Indeed, we measure the phase-front of the generated beam by performing the Fourier transform of the electron wavefunction at the exit facet of the hologram (see SM for more details~\cite{SM}). In order to do this, the thickness map of the hologram is obtained from the ratio between the elastic image and unfiltered image (see Ref.~\cite{grillo:14} for more details), and thus yields $t(x,y)$. This thickness $t(x,y)$ is directly related to phase modulation $\Delta\chi(x,y)$ introduced by the hologram, see Eq.~(\ref{eq:phase}). This was verified by simulating the beam diffraction from such a PM. As is shown in Fig.~\ref{fig:fig2}-({\bf a}), the simulation is in excellent agreement with the observed diffraction pattern, and that of the hologram with excluded dead region at the origin. The OAM spectrum is evaluated by transforming the electron wavefunction into polar coordinates and performing the Fourier transform in the azimuth coordinate. Figure~\ref{fig:fig3}-({\bf a}) shows the unraveled phase of the electron in hue-color, in polar coordinates. Thus, as expected by inspection of the shape of the diffracted beams, in both cases the OAM spectrum of the generated beam at the first-order of diffraction is delocalized. It is worth noticing that the generated beam is not simply a superposition of the two OAM values, $200\hbar$ and $400\hbar$, since these beams have different propagation axis. The OAM spectrum for both holograms are shown in Fig.~\ref{fig:fig3}-({\bf b}) and ({\bf c}), respectively. The peak of OAM values is located around, but not exactly at $200\hbar$. The reason lies in the fact that the high spatial frequencies can be mainly found in the missing details at the centre of the hologram. This means that the intensity of the beam, due to overlapping with the residual of the second-order of diffraction, is no longer azimuthally uniform as is expected for a beam that is an eigenstate of OAM.

%
\begin{figure}[b]
   \centering
   \includegraphics[width=0.38\textwidth]{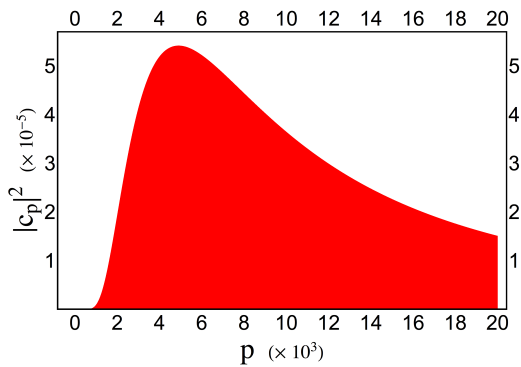} 
   \caption{(Online color) Radial index distribution of the generated highly-twisted EB with $m=200$ from the pitch-fork hologram shown in Fig.~\ref{fig:fig1}. For simplicity, we assumed a hologram with \emph{unlimited resolution}, and the size of hologram is reasonably larger than that of the impinging beam. The generated beam is delocalized in $p$ \qo{space} and has a peak around $p=4900$.}
   \label{fig:fig4}
\end{figure}
%

This can be explained by the decomposition of the large vortex in many isolated first-order singularities. To demonstrate this we need to characterize the position of the singularities. The topological charge $m$ about a circle $c$ is defined by $m=1/(2\pi)\oint_{c}\nabla\Phi\cdot d\mathbf{\ell}$, which gives the variation of phase around a path surrounding the origin. Using this definition we construct many infinitesimal \emph{steps} and map the position of the singularities. The resulting map is shown in Fig.~\ref{fig:fig2}-({\bf b}). However, in our analysis the contribution of the zeroth order is excluded, since an annular post-selection can remove its contribution. Finally, it is worth mentioning that the radial-index $p$ distribution of the EVBs generated by such a PM is different from those of the Laguerre-Gaussian (LG) modes. EBs with LG mode distribution have a well-defined radial index $p$, and are shape-invariant upon free-space propagation. This is particularly important, since the evolution of EVBs in the presence of a longitudinal magnetic field $B$ forms non-diffracting LG beams, so-called Landau states~\cite{bliokh:12}. It has been shown that the transverse energy ${\cal E}_\bot$ of these beams can be quantized according to Landau levels~\cite{schattschneider:14}, given by ${\cal E}_{\bot}=\hbar\Omega(2p+m+|m|+1)$, where $\Omega=eB/(2m_e)$ is the Larmor frequency corresponding to the g-factor of 1. The quantized transverse energy of the electron depends on the radial index of the beam. The radial distribution of the EB at a given propagation distance $z$ from the PM is given by $f(r,z)\,\exp{(im\phi)}$, with
%
\begin{align}\label{eq:radial}
	f(r,z)\propto\frac{e^{-\frac{i\pi}{z\lambda}r^2}}{z\lambda}\int_{r_\text{min}}^{r_\text{max}}\,J_{m}\left(\frac{2\pi r r'}{z\lambda}\right)\,e^{-\left(1+\frac{i\pi w_0^2}{z\lambda}\right)\left(\frac{r'}{w_0}\right)^2}\,r' dr',
\end{align}
where $J_m(x)$ is the Bessel function of integer order $n$ and $w_0$ is the beam radius at the PM, respectively. $r_\text{min}$ and $r_\text{max}$ are given by the active region of the hologram. For instance, for a relatively wide hologram of the first kind (the hologram with the central dead region) $r_\text{min}=0$ and $r_\text{max}\rightarrow\infty$. Thus, the emerging beam from the hologram is given in terms of Hypergeometric-Gauss (HyGG) beams~\cite{karimi:07}~(see SM for more details~\cite{SM}). Figure~\ref{fig:fig4} shows the radial-index distribution of the generated beam from an $m=200$ pitch-fork hologram. For the hologram with a truncated central region the radial distribution of the electron is more complicated and should be calculated by Eq.~\ref{eq:radial}. Therefore, the generated EB is a superposition of different LG modes having an azimuthal winding index $m$ and radial index $p$ distributions shown in Fig.~\ref{fig:fig3} and Fig.~\ref{fig:fig4}, respectively. These imply a non-monochromatic transverse energy distribution due to the indetermination of both, in the OAM and radial quantum indices.

In summary, we generated a highly-twisted EB with an OAM quanta of $200\hbar$. This beam is generated by a holographic approach in which a nano-fabricated pure phase hologram is used to manipulate a conventional beam into a form of EVB. The spectrum of OAM is calculated by measuring the thickness of the hologram and elaborating the diffraction pattern. Our analysis reveals that due to limited details on the PM, especially close to the origin, the generated beam contains a distribution of OAM. We believe that holographic generation of a \emph{high}-$m$ vortex beam can potentially be an interesting method to explore the transition between the quantum and classical regimes of electromagnetic radiations inside a medium. Thus, it may be used to exploit the magnetic property of materials such as magnetic dichroism.

E.K. and R.W.B. acknowledge the support of the Canada Excellence Research Chairs (CERC) Program.


\clearpage
\onecolumngrid
\setcounter{figure}{0} \renewcommand{\thefigure}{S\arabic{figure}}
\setcounter{section}{0} \renewcommand{\thesection}{S\arabic{section}}
\setcounter{equation}{0} \renewcommand{\theequation}{S\arabic{equation}}
\section*{Supplementary Material for ``Holographic generation of highly twisted electron beams''}
\section{Theory}
\subsection{Computer Generated Hologram}
As discussed in the main text, we fabricate a pure-phase hologram by milling a silicon-nitride membrane with a Ga-ion focused ion beam (FIB). This pure-phase mask is used to tailor the transverse state of the electron wavefunction. When an electron traverses through such a mask, neglecting a small absorption, its wavefunction is modified in-homogeneously in the transverse plane. Such a modification depends on the thickness of the mask $t(x,y)$, which varies in the transverse plane. In order to generate an electron beam which possesses a well-defined value of twist (orbital angular momentum) along propagation direction, the transverse thickness must vary according to $\alpha(x,y)=m\,\phi+2\pi x/\Lambda$, where $\Lambda$ is the grating period and $\phi=\arctan{(y/x)}$ with $x$ and $y$ being cartesian coordinates. A simple example of this variation is a sinusoidal function $t(x,y)=t_0\left(1+\cos{\alpha\left(x,y\right)}\right)/2$, which gives a symmetric wavefunction for the electron -- here $t_0$ is the modulation depth. In an electron microscope the transverse distribution for electron wavepacket is Gaussian with a standard deviation of $w_0/2$, i.e. $\psi_\text{initial}(x,y)=N\,e^{-r^2/w_0^2}$, where $r=\sqrt{x^2+y^2}$ is the radius in the polar coordinate and $N$ is a normalization constant. The electron state after passing through such a phase-mask is then given by
\begin{align}\label{eqs:wavefunction}
	\psi_\text{final}(x,y)=\psi_\text{in}(x,y)\,e^{i\,t(x,y)}.
\end{align}
The transmission function can be expanded in terms of Bessel function of the first kind, i.e. $J_n(.)$, according to Jacobi--Anger expansion.Thus, we have
\begin{align}\label{eqs:wavefunction2}
	\psi_\text{final}(x,y)&=N\,e^{i\,t_0/2} e^{-r^2/w_0^2}\,e^{i\,({t_0}/{2}) \cos{\alpha\left(x,y\right)}}\cr
	&=N\,e^{i\,t_0/2} e^{-r^2/w_0^2}\,\sum_{n=-\infty}^{+\infty}i^n\,J_{n}({t_0}/{2})\,e^{i\,n\,\alpha\left(x,y\right)},
\end{align}
where $n$ is an integer number. As can be seen, the final wavefunction of the electron is delocalized in space and distributed in equiangular angles of $2\pi x/\Lambda$ along $x$ with a symmetric probability of $|J_n(t_0/2)|^2$. Let us consider the zero and the first-order of diffractions, i.e. $n=0$ and $n=1$. The wavefunction of the electron for these two order of diffractions are given by:
\begin{align}\label{eqs:wavefunction3}
	\psi_\text{final}^{0}(x,y)&=N\,e^{i\,t_0/2} e^{-r^2/w_0^2}\,J_{0}({t_0}/{2}),\cr
	\psi_\text{final}^{1}(x,y)&=N\,e^{i\,t_0/2} e^{-r^2/w_0^2}\,i\,J_{1}({t_0}/{2})\,e^{i\,\alpha\left(x,y\right)}.
\end{align}
Therefore, the first-order of diffraction possesses an additional phase that is given by $\alpha(x,y)=m\,\phi+2\pi x/\Lambda$. Thus, it carries an OAM value of $m$, and is spatially diffracted into a specific angle of $2\pi\,x\lambda/\Lambda$, where $\lambda$ is the de Broglie wavelength of the electron. On the other hand, the zeroth order of diffraction carries no OAM and has the same distribution as the input wavefunction.

\subsection{Radial Index of the Generated Electron Beam}
The generated beam at the first-order of diffraction of Eq.~(\ref{eqs:wavefunction3}) has a twisted phase-front and a Gaussian amplitude. One can rewrite this beam in the dimensionless coordinate of $(\rho=r/w_0,\phi,\zeta=z\,\lambda/(\pi\,w_0^2))$;
\begin{align}\label{eqs:wavefunction4}
	\psi_\text{final}^{1}(\rho,\phi)=N'\,e^{-\rho^2}\,e^{i\,m\,\phi}.
\end{align}
where $N'$ is a normalization constant. This beam is different from that of the Laguerre-Gauss (LG) ones, i.e. $\rho^{|m|}\,e^{-\rho^2}\,e^{i\,m\,\phi}$, since the parabolic term of $\rho^{|m|}$ is missing. It has already been shown in Ref.~\cite{skarimi:07} that such a beam is a solution to the paraxial wave equation and is a sub-family of Hypergeometric-Gauss beams. However, it is worth trying to expand this wavefunction in the basis of other complete solutions of the paraxial wave equation such as the LGs. This is particularly important, since LG modes are shape-invariant upon free-space propagation, and also represent eigenstates of other fundamental quantum interactions such as the quantum Hall effect~\cite{sschattschneider:14}. Therefore, we can expand Eq.~(\ref{eqs:wavefunction4}) in terms of LG modes;
\begin{figure}[b]
   \centering
   \includegraphics[width=0.8\textwidth]{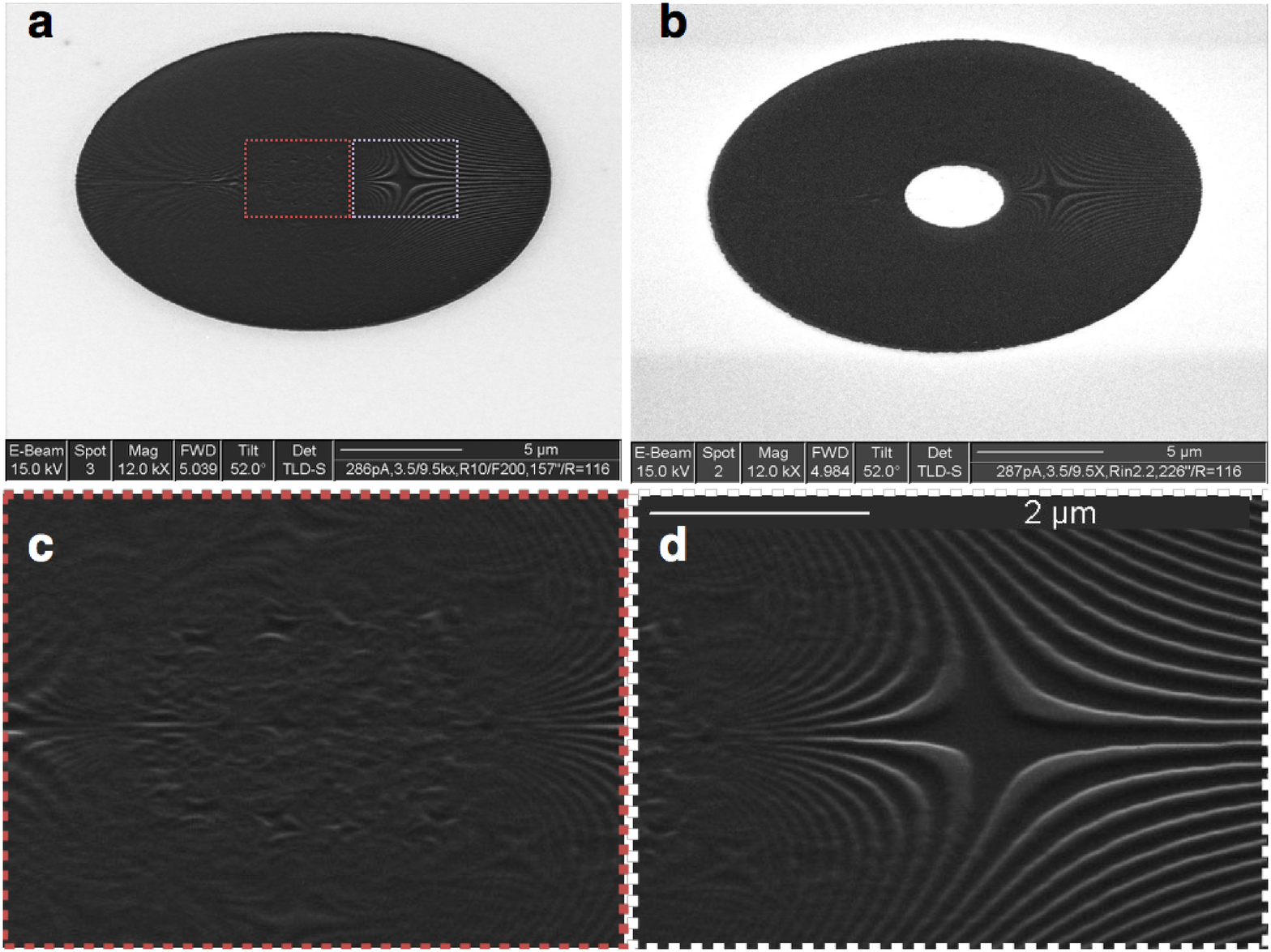} 
   \caption{(online color) Scanning electron microscope (SEM) image of the nano-fabricated computer-generated hologram (CGH) for generating an electron vortex beam with 200 twists. ({\bf a}) and ({\bf b}) are the fabricated CGHs with and without the dead region close to the origin. ({\bf c}) and ({\bf d}) show the zoomed-in regions indicated by dashed-red and dashed-white rectangles in ({\bf a}). As can be seen from fine structures in ({\bf c}), the hologram carrier changes more rapidly at the origin, thus the phase alteration goes beyond the spatial resolution of the hologram.}
   \label{figs:fig1}
\end{figure}
\begin{align}\label{eqs:expansion1}
	\psi_\text{final}^{1}(\rho,\phi)&=N'\,e^{-\rho^2}\,e^{i\,m\,\phi}\cr
	&=N'\sum_{p=0}^{+\infty} c_p\,\text{LG}_{p,m}(\rho,\phi),
\end{align}
where $c_p$ are the expansion coefficients,
$$\text{LG}_{p,m}(\rho,\phi)=\sqrt{\frac{2^{|m|+1}\,p!}{\pi\,(p+|m|)!}}\,\rho^{|m|}\,e^{-\rho^2}\,L_{p}^{|m|}\left(2\rho^2\right)\,e^{i\,m\,\phi}$$ are LG modes and $L_{p}^{|m|}(.)$ is the generalized Laguerre polynomial. A straightforward calculation shows that $c_p$ coefficients are given by 
\begin{align}\label{eqs:expansion2}
	c_p=\frac{|m|}{2}\,\frac{\Gamma(p+|m|/2)}{\sqrt{p!\left(p+|m|\right)!}},
\end{align}
where $\Gamma(.)$ is the gamma function. Therefore, the emerging beam from the CGH has a well-defined value of OAM, but its radial index is in a superposition state of an infinite number of the positive-integer $p$ with a weight given by Eq.~(\ref{eqs:expansion2}). Recall that the wavefunction of Eq.~(\ref{eqs:expansion1}) has an essential singularity at the origin, and well-defined radial \emph{local-minima} that changes upon propagation. This lies on the fact that different LG modes have different Gouy phases, thus relative phase between expansion terms in Eq.~(\ref{eqs:expansion1}) changes upon propagation and results a non-shape invariant distribution. 

\section{Experimental Results}
\subsection{Fabricated Holograms}
We implement our recent method reported in Ref.~\cite{sgrillo:14} for manufacturing the pure-phase mask. In this technique a dual-beam instrument (FEI Strata DB235M), combining a Ga-ion focused ion beam (FIB) and a scanning electron microscope (SEM) is used to pattern the CGHs by FIB milling on silicon-nitride membranes coated with a gold film. The nano-fabricated holograms have $20~\mu$m wide aperture and the desired patterns. Figure~\ref{figs:fig1}-({\bf a}) and ({\bf b}) show the scanning electron microscope (SEM) image of the fabricated holograms for holograms with and without the dead-zone central regions. The dead-zone region in Fig.~\ref{figs:fig1}-({\bf b}) is stopped by a gold film of $2~\mu$m, which was chosen based on observing the high-oscillation of the hologram carrier at the origin. Fine structure of these holograms are also shown in Fig.~\ref{figs:fig1}-({\bf c}) and ({\bf d}) for two different regions on the nano-fabricated holograms. As can be seen, these fine-structures are lost at the origin due to the effects that are discussed in the main text.

\begin{figure}[h]
   \centering
   \includegraphics[width=0.7\textwidth]{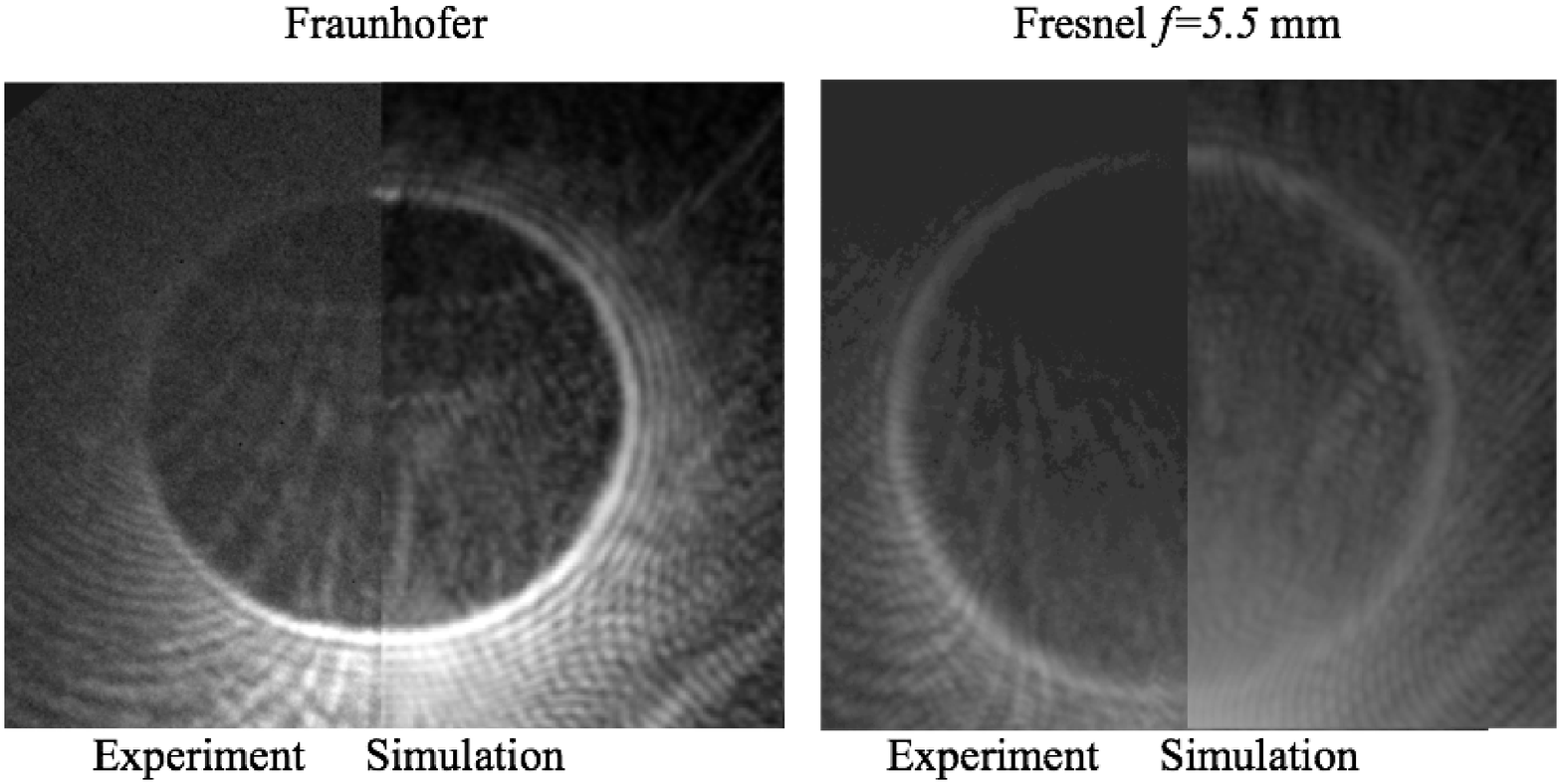} 
   \caption{Experimentally observed and thickness-map-based simulated distribution of the diffracted electron beams from nano-fabricated holograms. ({\bf a}) and ({\bf b}) show the electron distributions at two different propagation distance. ({\bf a}) corresponds to the Fraunhofer and ({\bf b}) corresponds to the Fresnel plane with a defocus of $f=5.5$~mm.}
   \label{figs:fig2}
\end{figure}
\subsection{Measuring the Phase-Front of the Diffracted Electron Beam}
In order to measure the phase-front of the diffracted electron beam, we acquired images at different defoci ranging from image- to diffraction-plane. The paradigm for this case would be to use the retrieval method like transport of intensity method (TIE). Under good coherence condition the inversion produces a  solution that is bound to describe the phase at every plane, and this solution is unique~\cite{slubk:13}. Rather than trying the inverse method, we can take advantage of the fact that we already have a test solution based on the \emph{thickness map measurement}. If the calculated image at different defoci matches the corresponding experimental image, we can be sure that the solution is correct. Figure~\ref{figs:fig2}-({\bf a}) and ({\bf b}) show two examples of the experimentally observed and simulated distribution based on thickness map measurement for electrons diffracted by our fabricated phase-mask upon propagation. Figure~\ref{figs:fig2}-({\bf a}) shows the electron distribution exactly at the Fraunhofer plane, and Fig.~\ref{figs:fig2}-({\bf b}) shows the distribution at the Fresnel plane with a defocus of $f=5.5$~mm.  The value of the defocus has been deduced by the distance between the transmitted and the centre of the diffraction. These have been obtained with no fitting parameters.  These images are recorded using Lorentz mode in JEOL machine under \qo{low mag} mode, where the main objective lens was turned off and the objective of the mini-lens below the sample was turned on. The spherical aberration of this lens is known to be on the order of $10$~cm and a strong defocalization takes from image to Fresnel and Fraunhofer plane. However only images at the focus are calibrated while diffraction is not calibrated. The Fraunhofer plane is obtained with a nominal defocalization of about $60$~mm.

%


\end{document}